# Numerical optimization of spherical VLS grating X-ray spectrometers

V.N. Strocov, T. Schmitt, U. Flechsig and L. Patthey

*Swiss Light Source, Paul Scherer Institut, CH-5232 Villigen-PSI, Switzerland*

G.S. Chiuzbăian

*UPMC Univ. Paris 06, CNRS UMR 7614, Laboratoire de Chimie Physique – Matière et Rayonnement, 75321 Paris Cedex 05, France*

**Abstract**

Operation of an X-ray spectrometer based on a spherical variable-line-spacing grating is analyzed using dedicated ray-tracing software allowing fast optimization of the grating parameters and spectrometer geometry. The analysis is illustrated with optical design of a model spectrometer to deliver a resolving power above 20400 at photon energy of 930 eV (Cu *L*-edge). With this energy taken as reference, the VLS coefficients are optimized to cancel the lineshape asymmetry (mostly from the coma aberrations) as well as minimize the symmetric aberration broadening at large grating illuminations, dramatically increasing the aberration-limited vertical acceptance of the spectrometer. For any energy away from the reference, we evaluate corrections to the entrance arm and light incidence angle on the grating to maintain the exactly symmetric lineshape. Furthermore, we evaluate operational modes when these corrections are coordinated to maintain either energy independent focal curve inclination or maximal aberration-limited spectrometer acceptance. The results are supported by analytical evaluation of the coma term of the optical path function. Our analysis gives thus a recipe to design a high-resolution spherical VLS grating spectrometer operating with negligible aberrations at large acceptance and over extended energy range.

**Keywords: Resonant inelastic X-ray scattering, X-ray optics, X-ray spectrometers, spherical VLS gratings**

# 1. Introduction

RIXS (Resonant Inelastic X-ray Scattering) is a synchrotron radiation based photon-in / photon-out spectroscopic technique, which gives information about charge-neutral low-energy excitations of the correlated electron system in solids, liquids and gases over the charge, orbital, spin and vibrational degrees of freedom (Kotani & Shin 2001). The recent progress in RIXS instrumentation (Ghiringhelli *et al*. 2006) allowing a resolving power $E/\Delta E$ better than 10000 has extended the RIXS experiment from the energy scale of charge transfer, crystal field and orbital excitations to that of magnetic and vibrational excitations (see, for example, Schlappa *et al*. 2009; Braicovich *et al*. 2010; Hennies *et al*. 2010).

The scientific progress in the field of RIXS is closely connected with progress in instrumentation. It pursues two main goals, improvement of the energy resolution towards progressively smaller energy scale of various charge-neutral excitations and, in view of low quantum yield of the RIXS process, improvement of the detection efficiency. In a variety of the optical schemes of RIXS spectrometers, the most popular are those based on a spherical grating (Fig. 1) as single optical element combining the dispersion and focusing actions. Although these instruments suffer from relatively small angular acceptance, their advantage is to deliver high energy resolution at downright simplicity. The first and still most widely spread high-resolution instrument of this type (Nordgren *et al*., 1989) uses a constant-line-spacing spherical grating. In order to cancel the coma aberrations, it operates in the Rowland circle geometry. This geometry is characterized by grazing inclination of the focal curve (FC) which results in necessity of large detector displacements with energy and, most important, small grazing angles of incidence on the detector incompatible with the modern directly illuminated CCD detectors. These disadvantages can be resolved with spherical variable-line-spacing (VLS) gratings (Osborn & Callcott, 1995; Cocco *et al*., 2004; Ghiringhelli *et al*., 1998 and 2006; Tokushima *et al*., 2006) which allow formation of any desired inclination of the focal plane towards upright as well as cancellation of the coma aberrations. The spherical VLS grating (SVLSG) is used, in particular, in the spectrometer SAXES (Ghiringhelli *et al.*, 2006) of the ADRESS beamline (Strocov *et al.*, 2010) at Swiss Light Source, Paul Scherrer Institut. This instrument delivers $E/\Delta E$ above 11000 at 1 keV photon energy, presently the highest achieved resolving power.

Optical design of the SVLSG based spectrometers is more complicated compared to the simple Rowland conditions and includes numerical computations. The optimal design should ensure minimal optical aberrations (and thus maximal resolution) at maximal angular acceptance (and thus spectrometer transmission). Here, we demonstrate the optical design of a model SVLSG spectrometer with $E/\Delta E$ above 20000. The grating parameters are optimized for a reference energy $E_{ref}$ of 930 eV (Cu *L*-edge important, for example, for the physics of correlated cuprates) to cancel the lineshape asymmetry coming mostly from the coma aberrations as well as to minimize the symmetric line broadening piling up at large illuminations. Furthermore, following our preliminary technical report (Strocov *et al.*, 2008), we evaluate adjustments of the spectrometer geometry upon variation of energy necessary to maintain the symmetric lineshape and constant focal curve inclination for any energy away from the reference.

## 2. Numerical procedure

Our evaluation of the grating parameters and spectrometer geometry described below used a dedicated software package TraceVLS written in MATLAB. The package is based on an effective numerical ray-tracing scheme devised to achieve maximal execution speed for further uses in optimization loops. Briefly, the ray-tracing is performed in two dimensions restricted by the dispersion plane of the spectrometer, shown schematically in Fig. 1. The rays from a point source are propagated towards the ideal spherical VLS grating. To deliver symmetric illumination of the grating relative to its center, the situation taking place when aligning the spectrometer in real experiment, the angular range of the rays is slightly asymmetric relative to the central ray. The rays diffract off the grating with the local groove density

$$a(\omega) = a_0 + a_1\omega + a_2\omega^2 + a_3\omega^3 + ...,\qquad(1)$$

where $\omega$ is the coordinate tangential to the grating surface in the center, according to the grating equation

$$\sin\alpha - \sin\beta = a_0 k\lambda,\qquad(2)$$

where $\lambda$ is the wavelength corresponding to the energy $E$, $k$ the diffraction order (positive for the internal), $\alpha$ the incidence angle on the grating and $\beta$ the diffracted beam angle (positive notation) relative to the surface normal. From the grating the rays propagate towards the detector whose position defined by the focal equation

$$r_2 = \frac{\cos^2\beta}{(\cos\alpha + \cos\beta)/R - (\cos^2\alpha)/r_1 + a_1 k\lambda},\qquad(3)$$

where $r_1$ and $r_2$ are the entrance and exit arms, respectively, and $R$ is the grating radius. The line profile is calculated as a histogram of the rays in the detector plane. This profile contains all optical aberrations such as the coma. Its further Gaussian broadening is due to the finite source size $\Delta_S$, grating slope errors $\Delta_{SE}$ and spatial resolution of the detector $\Delta_D$. Their contributions to the total linewidth are, respectively,

$$\Delta E_S = \Delta_S \frac{\cos\alpha}{r_1 a_0 k\lambda} E\qquad(4)$$

$$\Delta E_{SE} = \Delta_{SE} \frac{E}{\tan((\beta-\alpha)/2)}\qquad(5)$$

$$\Delta E_D = \Delta_D \sin\gamma \frac{\cos\beta}{r_2 a_0 k\lambda} E,\qquad(6)$$

hereinafter all widths being FWHM. We note in passing that our expression (5) is equivalent to its known form $\Delta E_{SE} = \Delta_{SE} \frac{(\cos\alpha + \cos\beta)E}{a_0 k\lambda}$ (Howells, 2001) where the denominator can be replaced according to the grating equation (2) and the trigonometric functions sum (difference) appearing in the nominator (denominator) is transformed to their product. The angle $\gamma$ in the expression (6) is the detector inclination relative to the central ray, which allows improvement of $\Delta E_\text{D}$. The total Gaussian line broadening is then simulated by convolution of the ray-tracing calculated (bare) line profile with a Gaussian whose width $\Delta E_\text{G}$ is taken as the vector sum $\Delta E_G = \sqrt{(\Delta E_S)^2 + (\Delta E_{SE})^2 + (\Delta E_D)^2}$. By virtue of this simplified computational method and extensive vectorization of the MATLAB code, a ray-tracing run with TraceVLS for a given set of parameters with a few thousands of rays takes less than a tenth of second on a low-end PC. Note that due to involving only the rays in the dispersion plane this procedure omits the 'smiley' line distortion in the perpendicular direction (see, for example, Tokushima *et al.*, 2006) which is however normally compensated by post-processing of the data.

The TraceVLS package was further used to optimize the grating parameters to deliver the narrowest symmetric profile at $E_\text{ref}$ (Sec. 3) as well as to adjust the spectrometer geometry to keep such profile when going away from $E_\text{ref}$ (Sec. 4). The principal obtained results were verified with generic ray-tracing codes PHASE (Bahrdt *et al.*, 1995) and RAY (Schäfers, 2008). The popular code SHADOW (available at http://www.nanotech.wisc.edu/shadow/) returns identical results starting from the year 2010 release which has fixed a bug on treatment of SVLSGs.

**3. Optimization of the grating parameters at reference energy**

Basics of the optical design procedure for SVLSG spectrometers are described, for example, in Ghiringhelli *et al.*, 2006. Here we follow a somewhat different route. We start with a definition of the following parameters: $E_\text{ref}$, $a_0$, $k$, $\alpha$, $\Delta_\text{S}$ and $\Delta_\text{D}$ introduced above, the total spectrometer length $L$ and the FC inclination angle $\gamma$ to match the optimal detector inclination angle. Then the $r_1$ and $r_2$ entrance and exit arm lengths are obtained by minimization of $\Delta E_\text{G}$ under the constraint $r_1 + r_2 = L$.

With $\Delta E_{SE}$ being independent of $r_1$ and $r_2$, this is equivalent to minimization of $(\Delta E_S)^2 + (\Delta E_D)^2$, where $\Delta E_S$ (4) decreases with $r_1$ and $\Delta E_D$ (6) increases with $r_1 = L - r_2$. Equating the derivative of this sum with respect to $r_1$ to zero takes us to the condition

$$r_1 = \frac{L}{\left(\dfrac{\Delta_D \sin\gamma \cos\beta}{\Delta_S \cos\alpha}\right)^{2/3} + 1} \qquad (7)$$

Compared to the seemingly obvious condition of balance of the $\Delta E_S$ and $\Delta E_D$ contributions (Ghiringhelli et al., 2006) the condition (7) improves the total $\Delta E_G$ (in our case by ~3000 in $E/\Delta E$) and displaces the grating towards the detector, requiring larger grating length for the same vertical acceptance of the spectrometer. The grating radius $R$ and the linear VLS term $a_1$ are then calculated as the analytical solutions of a system of two equations, which are the condition (3) on the focus to be at $r_2$ plus the condition imposed on the FC inclination

$$\tan\gamma = \frac{\cos\beta}{2\sin\beta - r_2\left((\tan\beta)/R + a_1/a_0\right)}. \qquad (8)$$

It should be noted that the possibility to control the FC inclination is an important advantage of the SVLSG spectrometers over the plane VLS ones.

For our model spectrometer, we have accepted realistic parameters of $E_0 = 930$ eV, $a_0 = 3500$ line/mm, $k = 1$ (internal), $\alpha = 88°$, $\Delta_S = 2$ µm, $\Delta_{SE} = 0.47$ µrad (corresponding to 0.2 µrad rms which is the present technological limit for spherical optics), $\Delta_D = 24$ µm, $\gamma = 20°$ and $L = 5000$ mm. The above procedure yielded $r_1 = 798.7$ mm, $r_2 = 4201.3$ mm, $R = 43241$ mm and $a_1 = 0.6377$ 1/mm². Ray-tracing calculations with TraceVLS, performed with the above parameters and a realistic grating illumination of 120 mm, yielded the results shown in Fig. 2 (a) as the bare line profile as well as the Gaussian broadened one. The profile is highly asymmetric due to aberrations dominated by the (primary) coma. With the Gaussian linewidth broadening $\Delta E_G = 45.4$ meV in our case, the aberrations deteriorate the spectrometer resolution to 84.8 meV.

The line asymmetry can be corrected by the $a_2$ coefficient of the VLS expansion. First, we should try to cancel the coma aberration predominantly contributing to the asymmetry. Evaluation of the optical path function (Howells, 2001; Peatman, 1997) and setting the $F_{30}$ (primary coma) term of its Maclaurin expansion to zero yields the condition

$$\frac{\sin\alpha}{2r_1}\left(\frac{\cos^2\alpha}{r_1}-\frac{\cos\alpha}{R}\right)-\frac{\sin\beta}{2r_2}\left(\frac{\cos^2\beta}{r_2}-\frac{\cos\beta}{R}\right)+\frac{1}{3}a_2 k\lambda = 0 \tag{9}$$

which allows analytical calculation of $a_2$ to cancel the coma. In our case it yields $a_2 = -0.975 \times 10^{-3}$ 1/mm$^3$. The results of ray-tracing performed with this $a_2$ at the 120 mm illumination are shown in Fig. 2 (*b, dotted lines*). The line asymmetry is greatly reduced, remaining only in some asymmetry of its foot.

However, the applicability of the analytical coma-free condition (9) is limited only to the coma aberration term and vicinity of the central ray, where the optical path function is derived. A numerical procedure should be applied to optimize $a_2$ taking into account the asymmetric aberrations of all orders as well as realistic grating illuminations. We used the TraceVLS ray-tracing procedure in an optimization loop to determine $a_2$ delivering the symmetric line profile as identified in the strict mathematical sense of zero skewness of the histogram. The optimized $a_2$ is obviously somewhat illumination dependent, but in practice the value found for large illuminations ensures that the line asymmetry stays negligible also with small illuminations, because all aberrations scale down with a power of 2 or stronger. In our case we performed the optimization with the above 120 mm illumination, which has returned $a_2 = -0.995 \times 10^{-3}$ 1/mm$^3$. The results of ray-tracing with this $a_2$ in Fig. 2 (*b, solid lines*) show a perfectly symmetric profile. Strictly speaking, this does not ensure that all asymmetric high-order aberrations vanish, but combine in a symmetric profile. We have checked that in the limit of vanishing illumination our optimization procedure returned the $a_2$ value identical within the numerical accuracy to the above analytical coma-free one. It should be noted that the difference between the analytical and optimized $a_2$ is only ~2%, well within the practical manufacturing accuracy. Interestingly, the analytical formula for $a_2$ from (Osborn & Callcott, 1995) returned a notably different value of -4.82 x 10$^{-3}$ 1/mm$^3$ yielding an asymmetric profile for all illuminations. While our optimization procedure allows full cancellation of the line asymmetry, the profile in Fig. 2 (*b*) still shows notable symmetric broadening and a broad foot due to higher-order aberrations piling up at large illuminations. In our case this deteriorates the spectrometer resolution from the $\Delta E_G$ = 45.4 meV Gaussian limit to 60.0 meV.

The remaining symmetric broadening can be reduced by optimization of the $a_3$ coefficient. Owing to a slight crosstalk of $a_3$ back to $a_2$ (in fact, separation of the line profile distortion into specific aberrations connected with particular $a_i$ coefficients is artificial and works only in vicinity of the central ray; their crosstalk increases with illumination) the optimization of $a_3$ with the highest accuracy should be performed under re-optimization of $a_2$ at each iteration step to keep the profile symmetric. For our model case with the 120 mm illumination this optimization returned $a_3$ = 2.02 x 10$^{-6}$ 1/mm$^4$ at almost the same $a_2$ = -0.986 x 10$^{-3}$ 1/mm$^3$. The corresponding ray-tracing calculations in Fig. 2 (*c*) show that the line profile has shrunk essentially to a delta-function (although with some structure on the meV scale) whose width is negligible compared to $\Delta E_G$. The spectrometer resolution has thus reached the Gaussian linewidth limit, delivering the resolving power $E/\Delta E$ = 20420. No attempt has been made to optimize VLS expansion coefficients higher than $a_3$ because they can hardly be realized with sufficient accuracy in a realistic manufacturing process.

The grating illumination is limited by increase of aberrations. In fact, this limit increases with $r_1$ in such a way that the corresponding vertical acceptance $\Delta\alpha$ stays roughly constant. In other words, the situations of small illumination of a grating close to the source and large illumination of a grating far from the source are roughly equivalent from the aberration point of view. We will therefore characterize the illumination by the corresponding $\Delta\alpha$ as a parameter more universal upon variations of $r_1$. The effect of $a_3$ on the aberration-limited spectrometer acceptance is illustrated in Fig. 3 which shows the total (aberration and Gaussian) linewidth as a function of $\Delta\alpha$ calculated without and with the optimized $a_3$. The low-aberration plateau, where the aberrations stay insignificant compared to the constant $\Delta E_G$, increases its width from ~2 to 7 mrad. The optimization of $a_3$ allows therefore operation of the spectrometer at much larger $\Delta\alpha$.

Compared to the presently most advanced spectrometer SAXES (Ghiringhelli *et al.*, 2006), the simulated spectrometer of the same dimensions promises an increase of $E/\Delta E$ by a factor of ~1.8 and the aberration-limited $\Delta\alpha$ by a factor of ~3.5. It should be noted that the spectrometer transmission can be further improved by another factor of ~3 by installing a collector mirror in the sagittal geometry in front of the grating to increase acceptance in the horizontal plane. Further increase of the angular acceptance may be achieved with optical schemes of Hettrick-Underwood (Hague *et al.*, 2005) or collimated-light plane grating (Agåker *et al.*, 2009) though compromising on resolution and transmission at higher soft-X-ray energies.

**4. Optimization of the spectrometer geometry for variable energy**

4.1. Lineshape dependence on the spectrometer geometry and angular acceptance

With the grating parameters optimized for certain $E_{\text{ref}}$, one can maintain the exactly symmetric line profile for any energy away from the reference by variation of the spectrometer geometry. We illustrate this in Fig. 4 (*top*) which shows, for our simulated spectrometer with the grating optimized for 930 eV, the ray-tracing calculated $\Delta E$ linewidth depending on $r_1$ and $\alpha$ for an energy of 530 eV (O *K*-edge). The illumination varies with $r_1$ and $\alpha$ over the plot to keep the $\Delta\alpha$ vertical acceptance constant and equal to 3 mrad (this value was chosen to stay within the low- aberration region, see below). It should be noted that these simulations always keep the spectrometer in focus, i.e. $r_2$ varies with $r_1$ and $\alpha$ over the plot according to the focal equation (3).

The resolution plot shows a prominent valley. Fig. 4 (*bottom*) illustrates evolution of the lineshapes upon crossing the valley by variation of $\alpha$ along the marked line of constant $r_1 = 600$ mm through the points *A*, *B* (bottom of the valley) and *C* separated by 0.05°. The point *B* is characterized by the symmetric lineshape (again, the asymmetric high-order aberrations may not exactly vanish in this point but yield a symmetric combination) whereas in the points *A* and *C* the asymmetry is already significant. Therefore, the best spectrometer resolution in the bottom of the valley corresponds exactly to the symmetric-profile (SP) lineshape. Note that upon crossing the valley the asymmetry tail flips from the left to the right side, which ensures there must exist a point where the asymmetry becomes exactly zero in the mathematical sense of zero skewness of the line profile. Therefore, the asymmetry can not merely be minimized, but totally *cancelled for any energy* away from the reference.

The resolution plot shows that the asymmetry cancellation can also be achieved by variation of $r_1$ for $\alpha = const$. Therefore, for every energy there are *two* alternative ways to maintain the SP spectrometer operation: either by optimizing the grating position along the beam to change $r_1$, or optimizing the pitch of the grating to change $\alpha$.

For $E_{ref}$ the resolution plot has the same pattern, i.e. the spectrometer can deliver the SP lineshape with $r_1$ and $\alpha$ different from the reference values (although with some increase of the symmetric aberration broadening and $\Delta E_G$ optimized for the reference geometry). This degree of freedom also allows compensation of certain manufacturing errors of the $a_2$ coefficient.

It is instructive to follow changes in resolution with increase of the $\Delta\alpha$ vertical acceptance. Fig. 5 shows the same resolution plot as in Fig. 4 but with $\Delta\alpha$ increased to 6 mrad which is beyond the low-aberration region. Similarly to the previous figure, the panels in the bottom illustrate evolution of the lineshapes upon crossing the valley along the $r_1$=600 mm line through the points *A*, *B* (bottom of the valley) and *C* separated by 0.05°. The valley bottom again corresponds to the SP-lineshape. With increase of $\Delta\alpha$ the valley becomes narrower, a consequence of the aberrations scaling up. This makes the spectrometer more sensitive to alignment. It is interesting to note a tiny bump appearing exactly in the bottom of the valley (i.e. the exactly symmetric profile has slightly larger FWHM) and a spike of FWHM piling up at the right border of the valley near the point *C* (due to formation of a double-peak structure in the line profile).

The effect of $\Delta\alpha$ is further illustrated in Fig. 6 which shows $\Delta E$ plots calculated for a series of $\Delta\alpha$. They correspond to two crossections of the above resolution plots, along the $r_1$=600 mm line as a function of $\alpha$ (*a*) and along the α=88.2 deg as a function of $r_1$ (*b*). As we have already seen in Fig. 5, with increase of $\Delta\alpha$ the valley narrows down, a spike of FWHM gradually forms on the right side of the valley, and a notable bump in the bottom piles up at large $\Delta\alpha$. A very slight displacement of the SP-point can be noted. Most important is however that the $\Delta E$ degradation in the SP-conditions stays insignificant, allowing the spectrometer operation at the highest transmission. Furthermore, the plot in Fig. 6 gives us an estimate of practical accuracy of the spectrometer settings. The curve for $\Delta\alpha$ = 4 mrad, for example, shows that if we accept a tolerance of 5% on degradation of $\Delta E$ relative to its minimum, the corresponding tolerances on $\alpha$ and $r_1$ are about ±0.02° and ±6 mm, respectively.

### 4.2 Evaluation of the symmetric-lineshape spectrometer settings

Corresponding to the resolution plot valley, the SP-trajectories in the $(r_1,\alpha)$ coordinates or the corresponding ones in the ($r_1,r_2$) coordinates define the spectrometer settings to maintain the SP-lineshape. We have calculated these trajectories for our model spectrometer in a range of energies from 430 to 1230 eV. First, we evaluated the SP-trajectories using the analytical coma-free condition (9). These "analytical" trajectories are displayed in Fig. 7 (*dotted lines*). Second, we used the TraceVLS ray-tracing procedure in an optimization loop similarly to the above determination of $a_2$. $\Delta\alpha$ in these calculations was kept at 5.2 mrad corresponding to the illumination used in the calculations at $E_{ref}$. These "numerical" trajectories are shown in Fig. 7 (*solid lines*). Obviously, the coma-free condition (9) gives an excellent approximation to the SP-trajectories. Nevertheless, the full ray-tracing analysis, taking into account the finite illumination and higher-order aberrations, introduces notable corrections, especially at the low-$r_1$ end. On average in the $r_1$ range displayed in the plot, the corrections are about 0.011° in $\alpha$ and 22 mm in $r_2$ resulting in increase of $E/\Delta E$ about 1130. The SP-trajectories in the ($r_1,\alpha$) and ($r_1,r_2$) coordinates, calculated in a range of energies, determine the required ranges of the $r_1$, $\alpha$ and $r_2$ mechanical motions.

It should be noted that prerequisite to maintain the SP-lineshape under energy variations is a mechanical flexibility of the SVLS spectrometer to vary at least two of the three parameters $r_1$, $\alpha$ and $r_2$. The beamline monochromators in general do not enjoy such a flexibility because of the fixed slit position. In (exactly focusing) spherical grating monochromators (Peatman, 1997) variation of $\alpha + \beta$ with the pre-mirror keeps the beam focused at the slit under energy variations, but there remain no degrees of freedom to cancel the line asymmetry away from $E_{ref}$ unless the grating is translated.

### 4.3. Fixed-Inclination and Maximal-Acceptance operation modes

For any energy one can achieve the SP spectrometer operation by setting different combinations of $r_1$ and $\alpha$ along the SP-trajectories. We will show that this remaining degree of freedom may be used in two ways, to maintain for each energy either fixed FC inclination angle $\gamma$ or minimal aberrations at large $\Delta\alpha$. We will refer to these two operation modes as the Fixed Inclination (FI) and Maximal-Acceptance (MA) modes.

To evaluate the FI mode, we have calculated the dependences of $\gamma$ defined by the equation (8) along the above 'numerical' SP-trajectories. Fig. 8 (*a*) displays these dependences as a function of $r_1$. They show dramatic variations and even jump from positive to negative values of $\gamma$, as seen for the lowest energy. It is not practical to follow these variations by changing the detector inclination angle, because this angle should normally stay around its optimal value chosen, on one side, as glancing as possible to reduce the effective pixel size and thus $\Delta E_\text{D}$ and, on another side, above the critical angle where the intensity starts to drop due to shadowing effects and increasing attenuation in the oxide dead layer (the 20º inclination angle adopted in our case is typical of the modern back-illuminated CCD chips).

Although relatively large focal depth of long spectrometers makes them not very critical on matching the FC to the detector inclination, one can find a mode to operate the spectrometer at fixed (energy independent) FC inclination. Indeed, Fig. 8 (*a*) shows that for any energy the SP-trajectories bear one point where the match is exact (crossings with the $\gamma = 20^\text{o}$ horizontal line). We have found the $r_1$ and corresponding $\alpha$ coordinates of these points by numerical solution of the equation (8) under the SP-constraint. The corresponding dependences of $\alpha$ and $r_1$ calculated in a wide energy range are shown in Fig. 8 (*b*). In this way, our analysis identifies the FI operation mode of the SVLSG spectrometer which maintains in a wide energy range the SP-lineshape *and* exact match of the FC to the detector inclination.

Principles of the MA mode are illustrated in Fig. 9 (*a*), which shows $\Delta E$ dependences of $r_1$ along the 'numerical' SP-trajectory from Fig. 7 calculated for an energy of 530 eV. Whereas for small $\Delta\alpha$ these dependences show a monotonous decrease of $\Delta E$ with $r_1$, for large $\Delta\alpha$ there develops a pronounced minimum at $r_1 \sim 750$ mm. In this point $\Delta E$ is almost independent of $\Delta\alpha$. Similarly to the effect at $E_{ref}$, this minimum appears due to the $a_3$ coefficient. Therefore, for any energy the SP-trajectories bear one point where the aberration-limited $\Delta\alpha$ is maximal, characteristic of the MA operation mode.

We have determined the $r_1$ and corresponding $\alpha$ coordinates of the MA points in an extended energy range by numerical minimization of $\Delta E$ under the SP-constraint. The results are shown in Fig. 9 (*b*). They identify the MA operation mode of the SVLSG spectrometer which maintains in a wide energy range the SP-lineshape *and* maximal aberration-limited $\Delta\alpha$.

Furthermore, we have investigated how large is the effect of $a_3$ to increase the aberration-limited $\Delta\alpha$ away from $E_{ref}$. The two bottom curves in Fig. 3 show the total linewidth at 530 eV as a function of $\Delta\alpha$ calculated in the MA mode with $a_3=0$ and with our optimized $a_3$. Although the optimization was performed at 930 eV, this $a_3$ increases the width of the low-aberration plateau from ~2 to 7 mrad, the effect as large as at $E_{ref}$.

Finally, we have compared in a wide energy range the FI and MA modes in terms of resolution. The calculations were performed with the $\Delta\alpha$ value of 5.2 mrad used in the calculations at $E_{ref}$. Fig. 10 (*solid lines*) shows the calculated $\Delta E$ dependences together with those of the Gaussian resolution limit $\Delta E_G$ (*dotted*). As expected, in the FI mode the $\Delta E$ values are generally above $\Delta E_G$ owing to the symmetric aberration broadening at this $\Delta\alpha$. The difference (predominantly due to the aberrations higher than coma) rapidly decreases with decrease of $\Delta\alpha$ and vanishes in the $\Delta\alpha = 0$ limit. The two dependences coincide at $E_{ref}$ where the $a_3$ coefficient was optimized to minimize this broadening. In the MA mode, by its design principle, the $\Delta E$ dependence almost coincides with its

$\Delta E_G$ limit, providing better resolution with large $\Delta\alpha$ compared to the FI mode. Note that the $\Delta E_G$ dependences are slightly different in the two modes due to different trajectories in the ($r_1$, $\alpha$, $r_2$) coordinates, see Figs. 9 and 10. Energy variations of the FC inclination in the MA mode, plotted in the corresponding panel of Fig. 10, are large.

Also shown in Fig. 10 (*dashed lines*) are the $\Delta E_S$ source size, $\Delta E_D$ detector and $\Delta E_{SE}$ slope error contributions to the total $\Delta E_G$. The resolution is limited almost purely by $\Delta E_D$. This demonstrates that improvement of the spatial resolution of X-ray detectors is the factor most important for further energy resolution progress of the soft-X-ray spectrometers.

### 4.4. Software tools

Based on the TraceVLS package, we have developed a user-friendly GUI-based program for fast determination of the optimal spectrometer geometry for varying energy, including the FI and MA modes. The GUI is shown in Fig. 11. First, in the box 'GRATING' one defines the grating parameters. Then in the box 'PARAMETERS' one defines the fixed spectrometer settings, including the detector inclination angle and some of the three geometry parameters $r_1$, $\alpha$ and $r_2$ necessary to calculate the remaining ones according to the focalization conditions defined in the box 'FOCUS MODE' below. If one checks the simple focus, the code calculates either $\alpha$ out of given ($r_1$, $r_2$) or $r_2$ out of given ($r_1$, $\alpha$) based on the focal equation (3). If one checks for the SP-focus, the code calculates ($\alpha$,$r_2$) out of given $r_1$ or ($r_1$,$\alpha$) out of given $r_2$ based on two conditions, the focal equation (3) plus zero asymmetry of the line profile as defined by numerical optimization. If one checks the focus in the FI or MA modes, the code calculates all three parameters ($r_1$,$\alpha$,$r_2$) based on the two above conditions plus the one that either $\gamma$ matches the detector inclination or the symmetric aberrations for given illumination are minimal, respectively. The results of calculations are displayed in the box 'RESULTS' as the calculated bare line profile, Gaussian broadening and the resulting total line profile, as well as numerical outputs such as various contributions to the total resolution and the diffraction angles.

Due to the fast ray-tracing scheme, the TraceVLS based GUI finds the optimal spectrometer settings for given energy in less than a second on a low-end PC for the simple or SP-focus, and a couple of seconds for the FI and MA modes. The user-friendly interface allows its use as an online tool in real experiment. It should be noted that similar optimizations using generic ray tracing software like SHADOW would be far more laborious owing to necessity to manually set up the computational parameters in each pass of the optimization loop. The code is written in MATLAB and is platform independent. It is available free for the academic users by writing to the first author.

## 5. Summary

We have analysed operation of a VLS-grating based X-ray spectrometer using a dedicated ray-tracing software package TraceVLS allowing fast optimization of the grating parameters and spectrometer geometry. The analysis is illustrated with optical design of a model spectrometer delivering $E/\Delta E$ above 20400 at a photon energy of 930 eV. With a reference energy $E_{\mathrm{ref}}$ chosen at 930 eV, the spectrometer geometry is evaluated to minimize the Gaussian line broadening due to the source size, grating slope errors and detector spatial resolution. The lineshape asymmetry (mostly due to the coma aberrations) is cancelled by optimization of the $a_2$ coefficient of the VLS power expansion. At small illuminations the obtained $a_2$ becomes identical to that yielded by the analytical coma-free condition derived from the optical path function. Furthermore, the remaining symmetric line broadening at large illuminations (due to higher-order aberrations) is reduced by optimization of $a_3$ which allows dramatic increase of the aberration-limited $\Delta\alpha$ acceptance of the spectrometer, in our case by a factor about 3.5. For any energy away from $E_{\mathrm{ref}}$, the exact asymmetry cancellation can be maintained by correcting either $r_1$ or $\alpha$. The corresponding SP-trajectories in the

($r_1$,α) and ($r_1$,$r_2$) coordinates are calculated from the analytical coma-free condition and, with better accuracy, by numerical minimization of the line asymmetry. The remaining degree of freedom to set different combinations of $r_1$ and α along the SP-trajectories is utilized to maintain either energy independent FC inclination (FI operational mode) or maximal aberration-limited Δα acceptance (MA mode) which exploits the effect of the $a_3$ coefficient to minimize the symmetric aberration broadening. In routine experimental work, the optimal $r_1$, α and $r_2$ spectrometer settings can be calculated in a fraction of second using our ray-tracing code wrapped in a user-friendly GUI. Our analysis gives thus a recipe to design and operate SVLSG spectrometers at large angular acceptance and in extended energy range without any notable degradation of resolution beyond the Gaussian broadening factors. These properties of the SVLS optical scheme along with its ultimate simplicity suggest its use in the $hv^2$ spectrometer (Strocov, 2010) where imaging and dispersion actions in two orthogonal planes are combined to deliver the full two-dimensional map of RIXS intensity with simultaneous detection in incoming and outgoing photon energies.

*Acknowledgments.* We thank G. Ghiringhelli, L. Braicovich, Y. Harada, R. Reininger and C. Hague for promoting discussions, and C. Quitmann and F. van der Veen for their continuous support of the RIXS project at Swiss Light Source.

**Figure captions**

Fig. 1. Scheme of the SVLSG spectrometer and the main notations.

Fig. 2. Effect of the $a_2$ and $a_3$ coefficients of the VLS expansion on the line profile for the model spectrometer at an illumination of 60 mm, calculated with (*a*) $a_2 = a_3 = 0$; (*b*) analytical coma-free $a_2$ (*dotted line*) and numerically optimized asymmetry-free $a_2$ (*solid*), with $a_3 = 0$; (*c*) numerically optimized $a_2$ and $a_3$. Shown are the bare line profiles (*red*) and the corresponding Gaussian broadened ones (*blue*). The profiles are normalized to the maximal amplitude. The optimization of $a_2$ delivers a symmetric profile, and $a_3$ suppresses the symmetric broadening at large illuminations.

Fig. 3. The total (aberration and Gaussian) linewidth $\Delta E$ depending on the $\Delta\alpha$ vertical acceptance without and with the $a_3$ coefficient (optimized at 930 eV) for energies of 930 eV (two upper curves) and 530 eV in the MA mode (two lower ones, see the text below). Optimization of $a_3$ dramatically increases the maximal illumination and thus aberration-limited $\Delta\alpha$ even away from $E_{\text{ref}}$.

Fig. 4. (*top*) Resolution as a function of $r_1$ and $\alpha$ calculated for an energy of 530 eV and $\Delta\alpha$ of 3 mrad. The bottom of the valley corresponds to exactly symmetric line profile. This is illustrated (*bottom*) by evolution of the lineshapes through the points *A*, *B* and *C* across the valley, calculated without (*red*) and with (*blue*) $\Delta E_G$ broadening.

Fig. 5. The same resolution plot as in Fig. 4 but with $\Delta\alpha$ increased to 6 mrad. The SP-valley narrows down, making the spectrometer more sensitive to alignment.

Fig. 6. Resolution plots as a function of $\alpha$ for $r_1 = 650$ mm (*a*) and as a function of $r_1$ for $\alpha = 88.2°$ (*b*) calculated with $\Delta\alpha$ increasing from 1 to 6 mrad in steps of 1 mrad. The SP-valley narrows down, but $\Delta E$ in its bottom increases only marginally.

Fig. 7. SP-trajectories in the $(r_1,\alpha)$ and $(r_1,r_2)$ coordinates calculated with $\Delta\alpha = 5.2$ mrad for energies going from 430 to 1230 eV in steps of 100 eV: 'Analytical' calculated from the coma-free condition (*doted lines*) and 'numerical' by ray-tracing based optimization (*solid*).

Fig. 8. (*a*) Variations of the FC inclination $\gamma$ along the 'numerical' SP-trajectories from Fig. 7; (*b*) Energy dependences of $\alpha$ and $r_1$ delivering the SP-lineshape at constant $\gamma = 20°$, identifying the FI operation mode.

Fig. 9. (*a*) $\Delta E$ dependences of $r_1$ along the 'numerical' SP-trajectory for 530 eV from Fig. 7 calculated with different $\Delta\alpha$. In their minimum $\Delta E$ is almost independent of $\Delta\alpha$, identifying the MA mode; (*b*) Energy dependences of $\alpha$ and $r_1$ for the MA mode.

Fig. 10. Energy dependences of the resolution $\Delta E$ and its Gaussian limit $\Delta E_G$ (*solid and dotted lines*) for the FI and MA modes. Also shown is the $\Delta E_G$ breakout into the source size $\Delta E_S$, detector $\Delta E_D$ and slope error $\Delta E_{SE}$ components (*dashed*) where $\Delta E_D$ dominates. The line marked $\gamma(E)$ in the MA panel shows energy variations of the FC inclination in this mode.

Fig. 11. Screenshot of the TraceVLS-based GUI to optimize the spectrometer geometry for different energies. With the grating optimized at 930 eV, the shown spectrometer settings deliver the SP-lineshape at 530 eV.

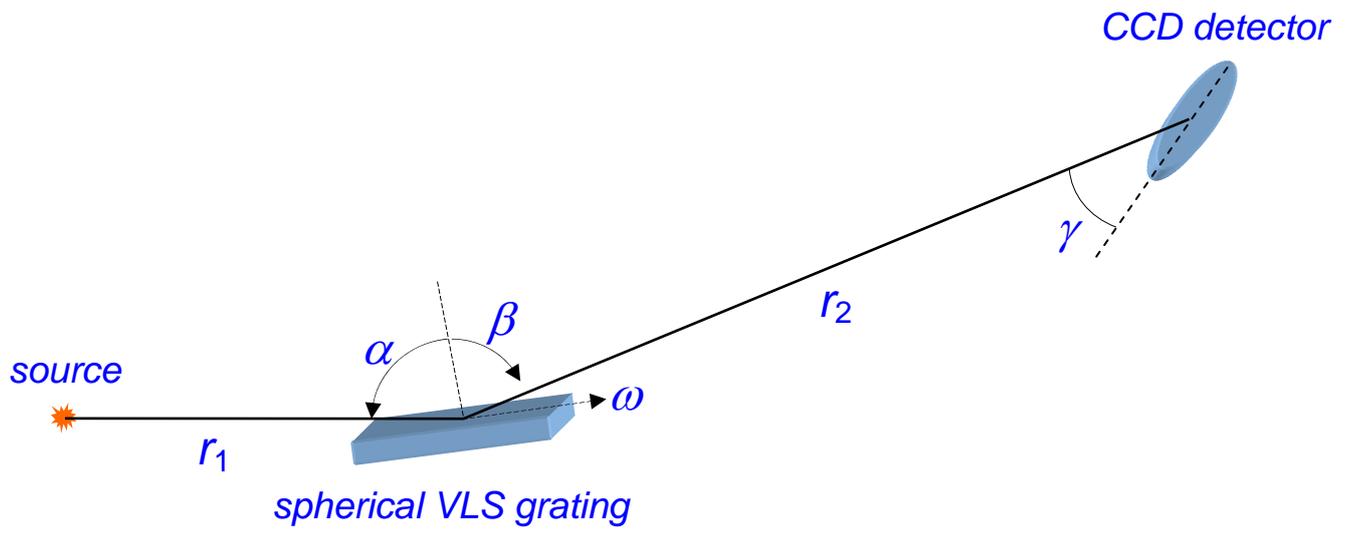

Fig. 1

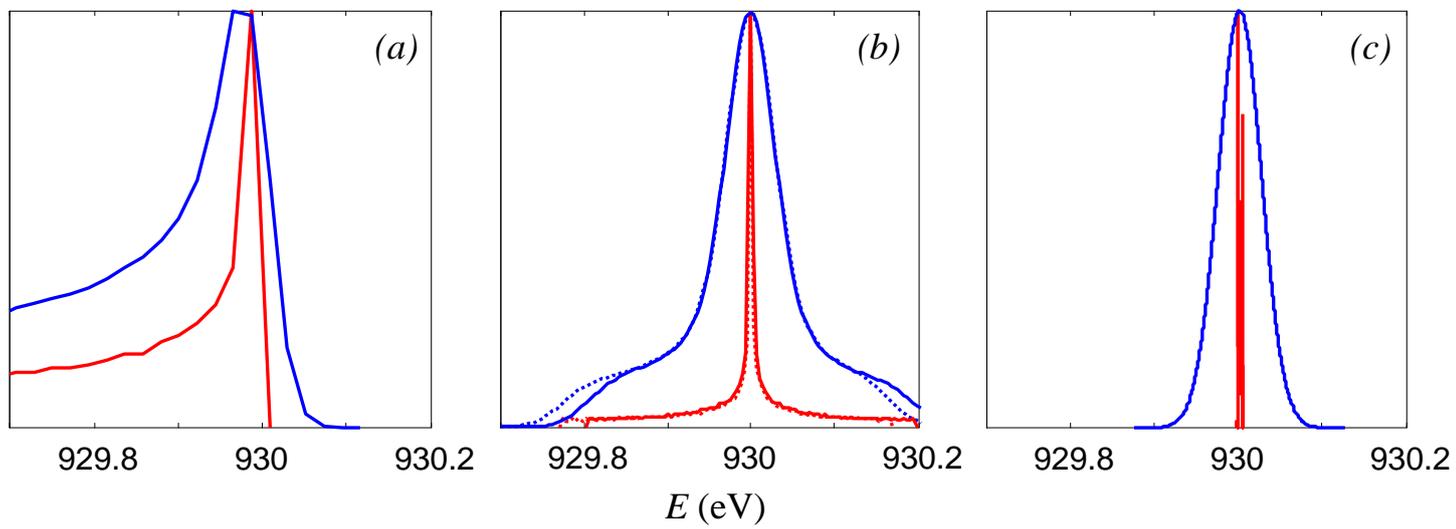

Fig. 2

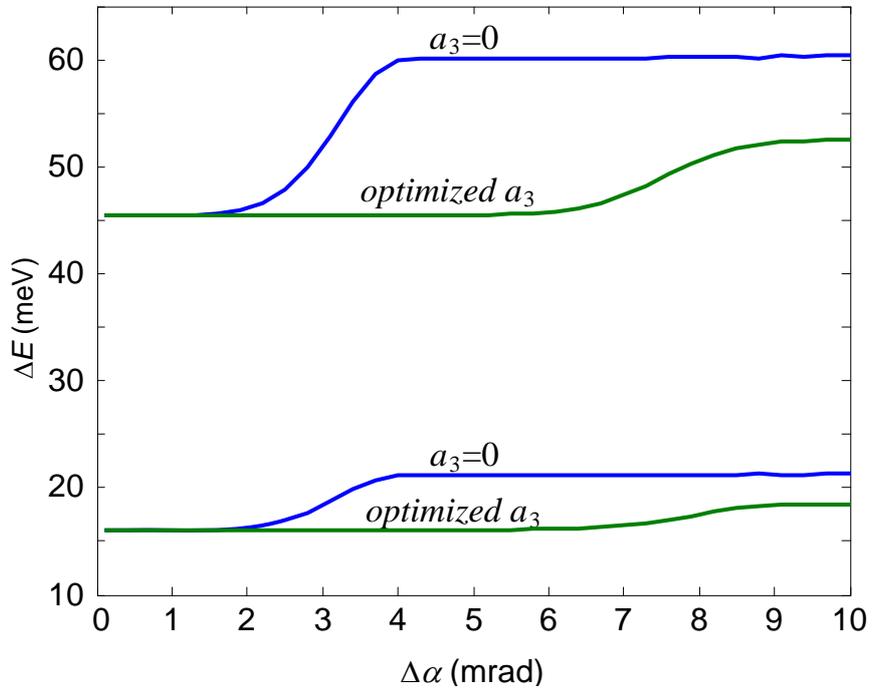

Fig. 3

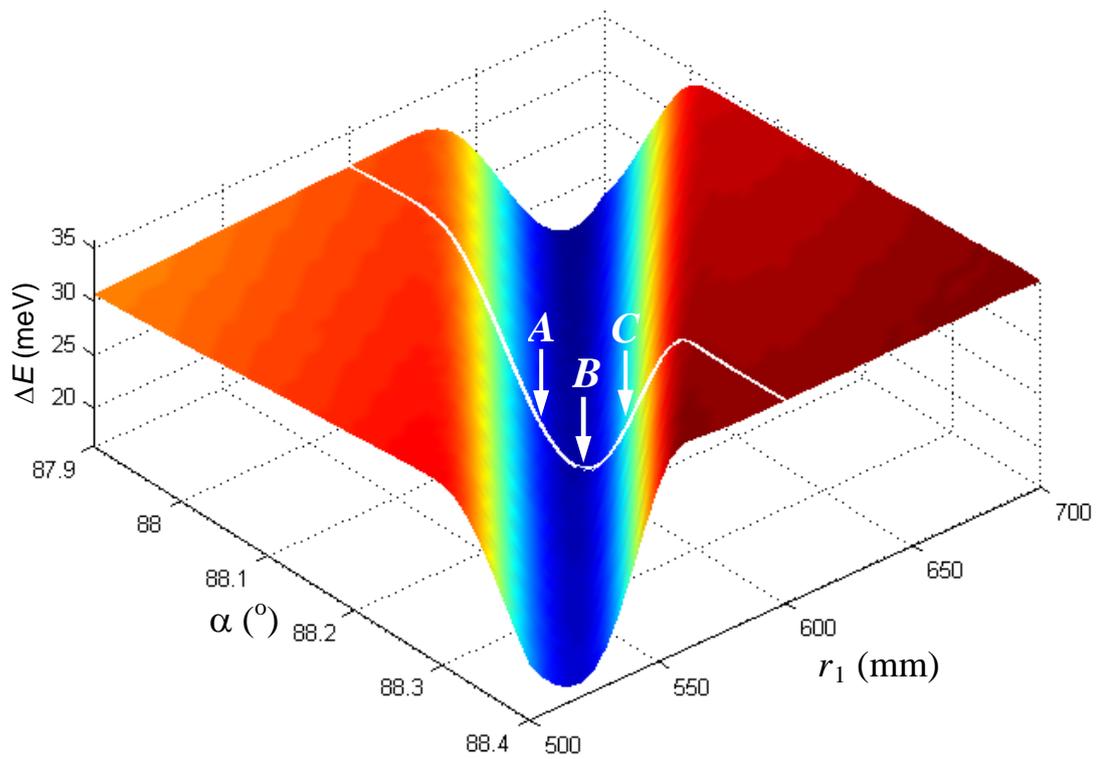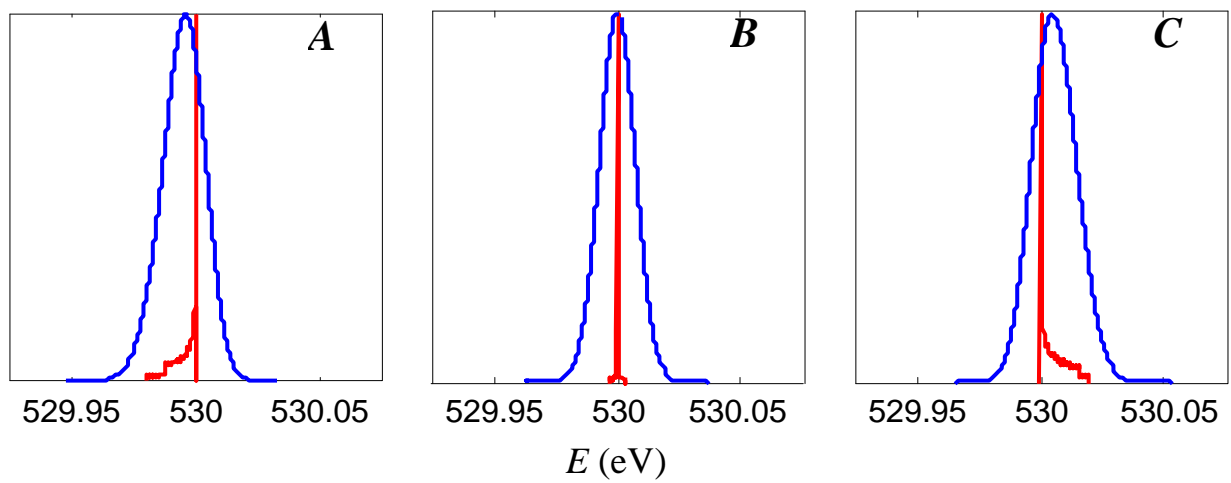

Fig. 4

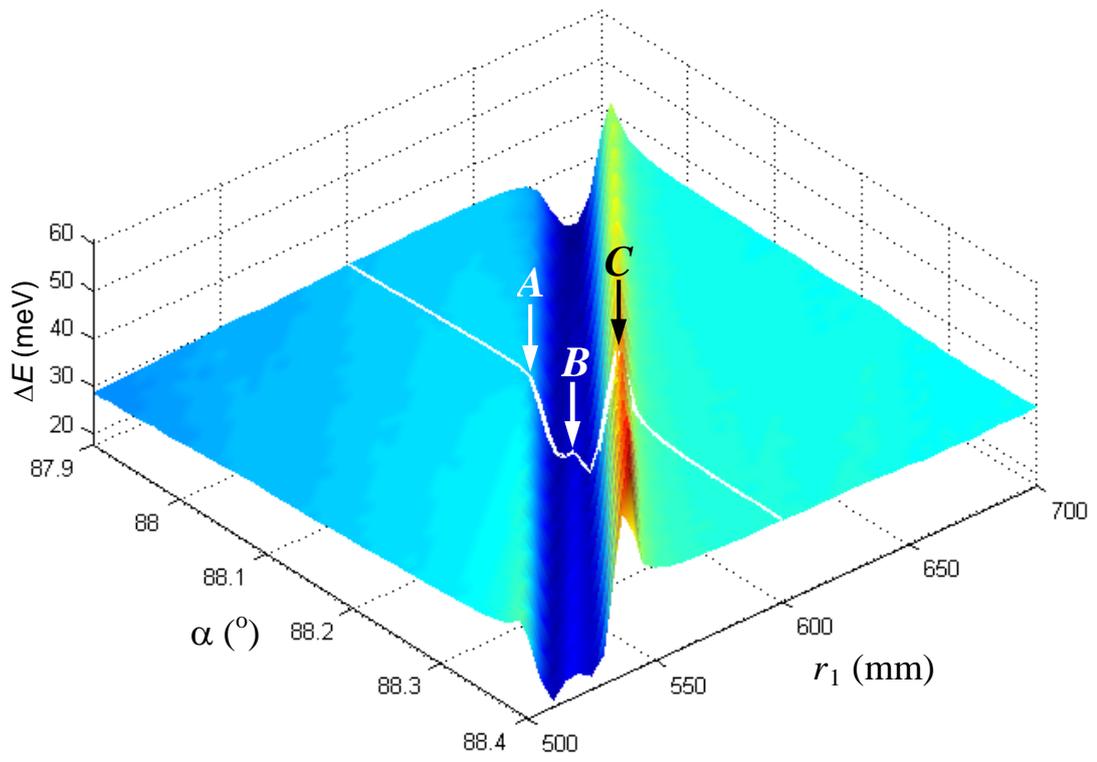
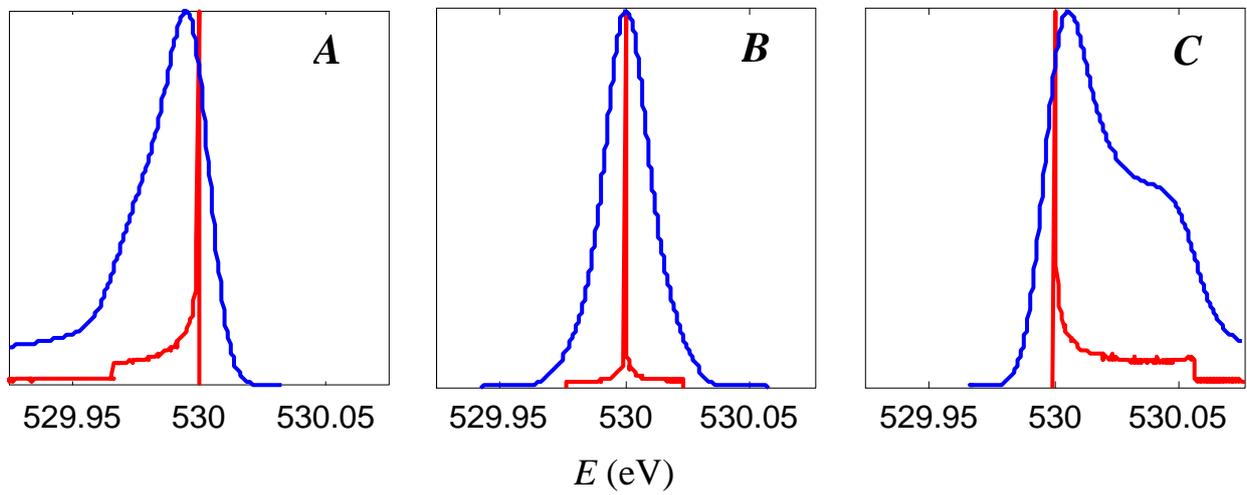

Fig. 5

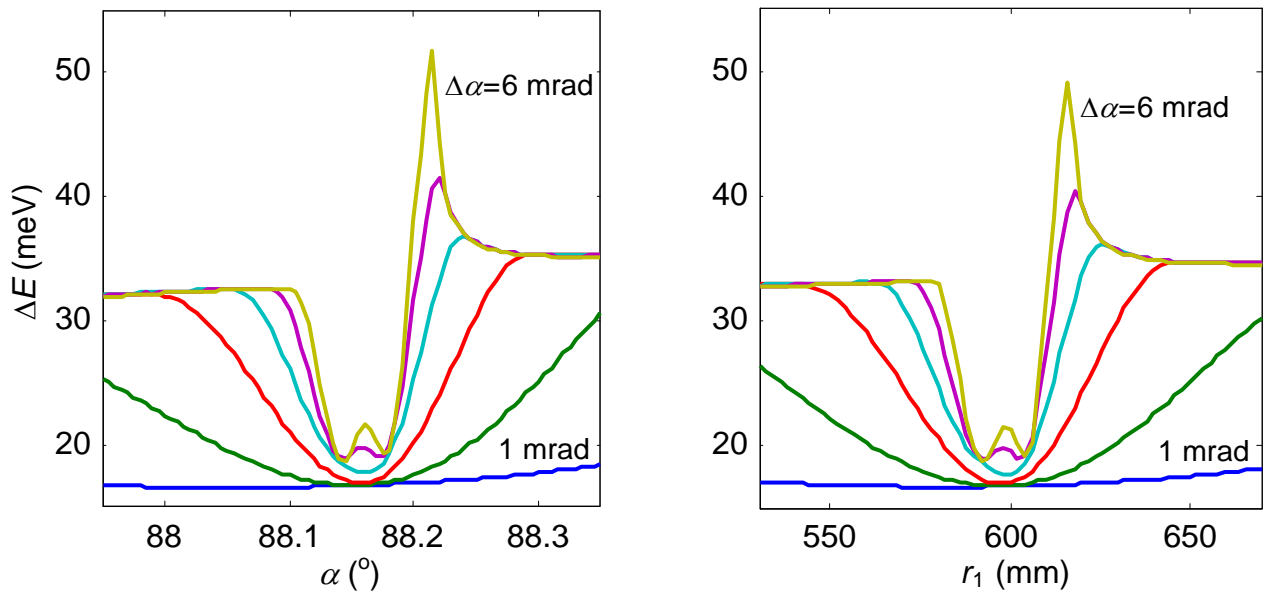

Fig. 6

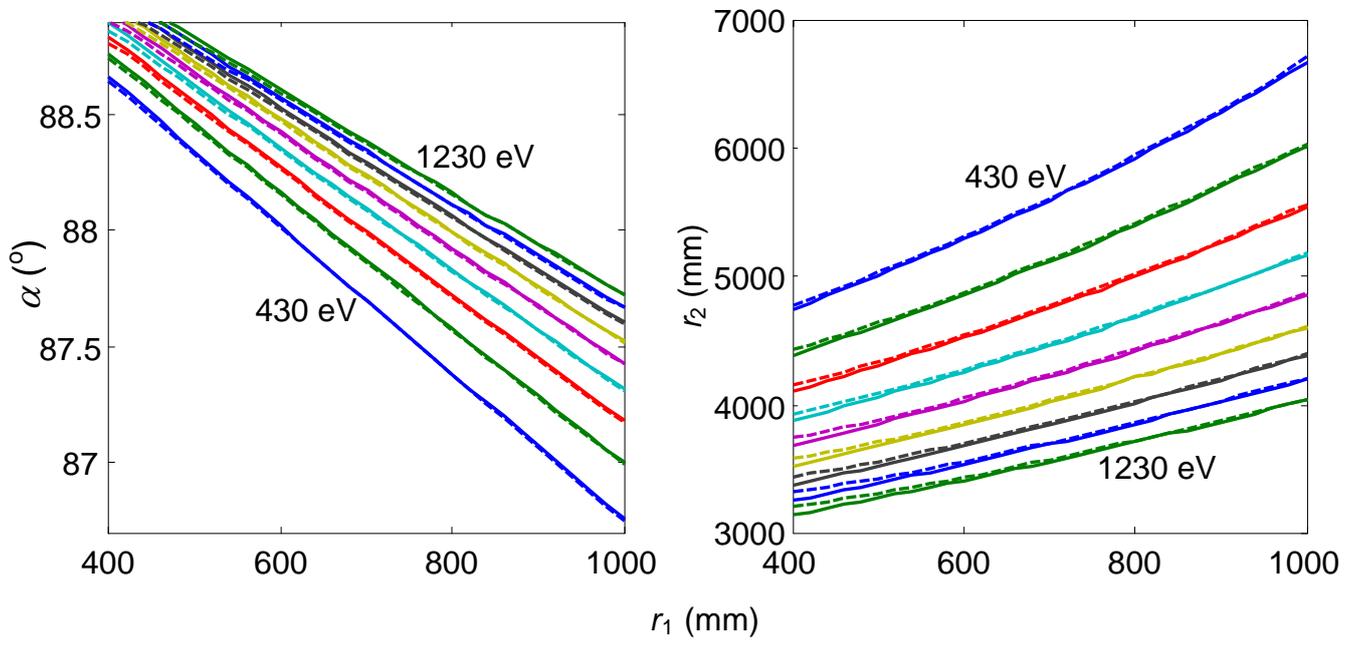

Fig. 7

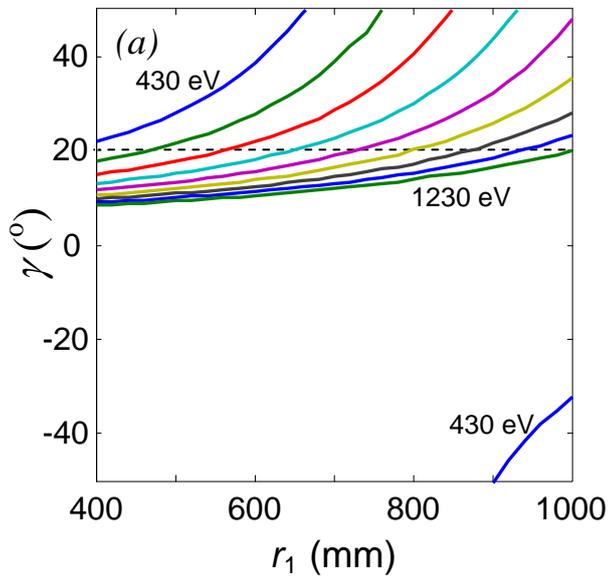 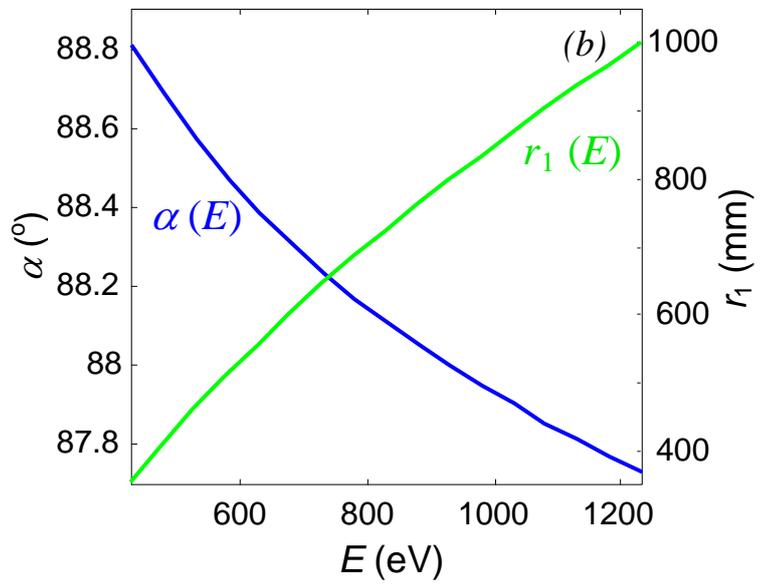

Fig. 8

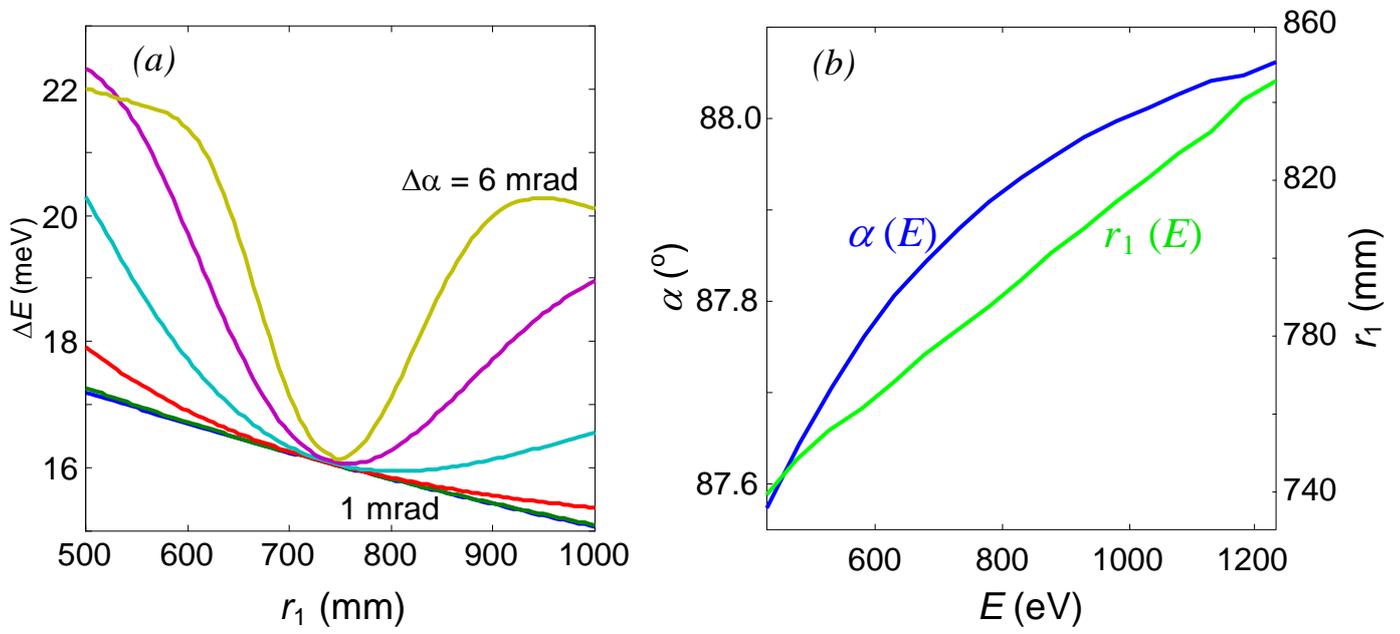

Fig. 9

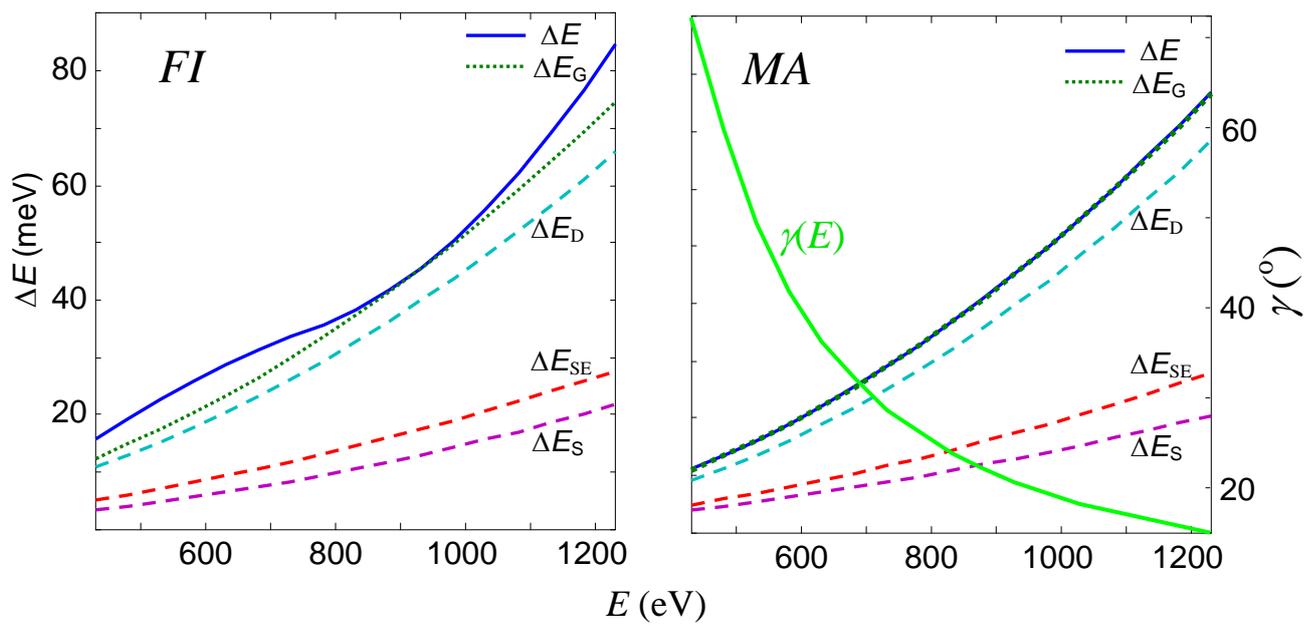

Fig. 10

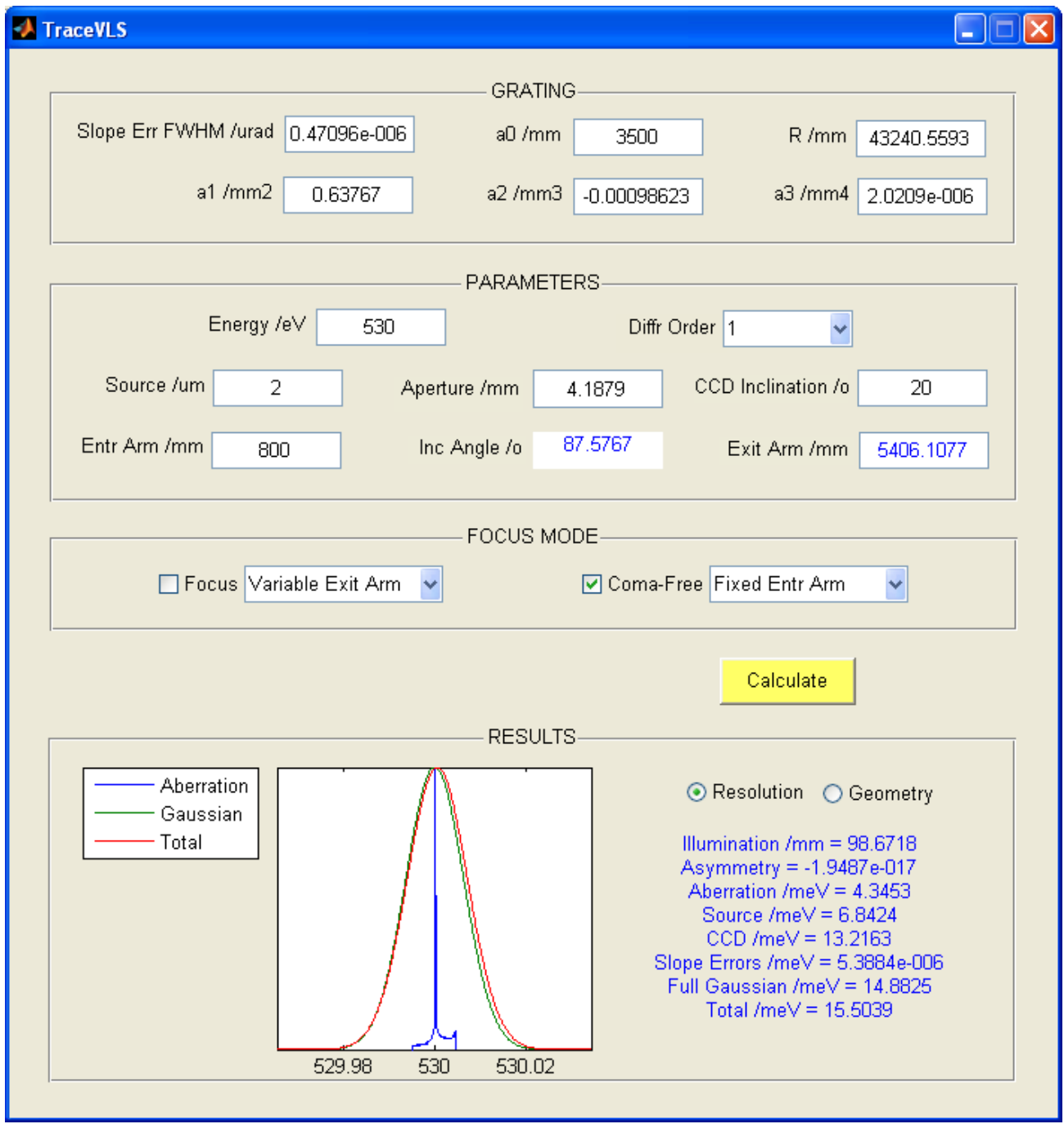

Fig. 11